\makeatletter
\newcommand{\dontusepackage}[2][]{%
  \@namedef{ver@#2.sty}{9999/12/31}%
  \@namedef{opt@#2.sty}{#1}}
\makeatother
\dontusepackage{subfigure}

\documentclass[]{sigchi}

\usepackage{lmodern}
\usepackage{amssymb,amsmath}
\usepackage{ifxetex,ifluatex}
\usepackage[usenames,dvipsnames]{color}
\usepackage{fixltx2e} 
\ifnum 0\ifxetex 1\fi\ifluatex 1\fi=0 
  \usepackage[utf8]{inputenc}
\else 
  \ifxetex
    \usepackage{mathspec}
    \usepackage{xltxtra,xunicode}
  \else
    \usepackage{fontspec}
  \fi
  \defaultfontfeatures{Mapping=tex-text,Scale=MatchLowercase}
  
\fi
\IfFileExists{upquote.sty}{\usepackage{upquote}}{}
\IfFileExists{microtype.sty}{%
\usepackage{microtype}
\UseMicrotypeSet[protrusion]{basicmath} 
}{}
\usepackage{listings}
\usepackage[font={small,bf},skip=2pt]{caption}

\def\plainauthor{Renaud Gervais, Jérémy Frey, Alexis Gay, Fabien Lotte and Martin Hachet}

\ifxetex
  \usepackage[setpagesize=false, 
              unicode=false, 
              xetex]{hyperref}
\else
  \usepackage[unicode=true]{hyperref}
\fi
\hypersetup{breaklinks=true,
            bookmarks=true,
            pdfauthor={\plainauthor},
            pdftitle={Introspectibles: Tangible Interaction to Foster Introspection},
            colorlinks=true,
            citecolor=black,
            urlcolor=blue,
            linkcolor=black,
            pdfborder={0 0 0}}
\usepackage{balance}

\title{Introspectibles: Tangible Interaction to Foster Introspection}

\numberofauthors{4}
\author{%
    \alignauthor{Renaud Gervais\\
        \affaddr{Inria, France}\\
        \email{renaud.gervais@inria.fr}}\\
    \alignauthor{Joan Sol Roo\\
        \affaddr{Inria, France}\\
        \email{joan-sol.roo@inria.fr}}\\
    \alignauthor{Jérémy Frey\\
        \affaddr{Univ. Bordeaux, France}\\
        \email{jeremy.frey@inria.fr}}\\
    \alignauthor{Martin Hachet\\
        \affaddr{Inria, France}\\
        \email{martin.hachet@inria.fr}}\\
  }

\date{}

\usepackage{times}

\usepackage{suffix}

\usepackage{subfig}

 \toappear{\emph{CHI'16}, May 07 - 12, 2016, San Jose, CA, USA \\
Copyright is held by the owner/author(s).}

\begin{document}

      \teaser{
      \includegraphics[width=\textwidth]{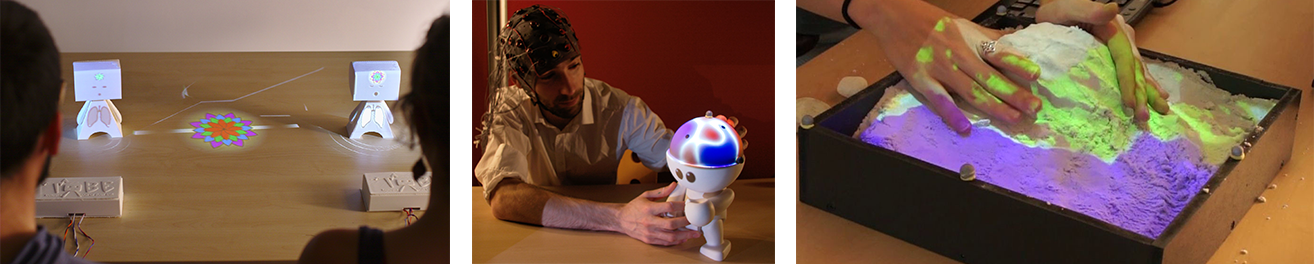}
      \caption{(Left) Tobe is a toolkit enabling the creation of tangible augmented
customized bio- and neurofeedback. Here, two users are relaxing together
by using Tobe to reach cardiac coherence. (Middle) Teegi is an augmented
tangible avatar that displays the user's brain activity in real time.
(Right) Inner Garden is an augmented sandbox, linked to the user's inner
states via physiological sensors, designed for contemplative play.\label{teaser}}
    }
  
\maketitle

\begin{abstract}
Digital devices are now ubiquitous and have the potential to be used to
support positive changes in human lives and promote psychological
well-being. This paper presents three interactive systems that we
created focusing on introspection activities, leveraging tangible
interaction and spatial augmented reality. More specifically, we
describe anthropomorphic augmented avatars that display the users' inner
states using physiological sensors. We also present a first prototype of
an augmented sandbox specifically dedicated to promoting mindfulness
activities.
\end{abstract}



\def \citep {\protect\cite}



\makeatletter
\def\url@leostyle{%
  \@ifundefined{selectfont}{\def\UrlFont{\sf}}{\def\UrlFont{\small\bf\ttfamily}}}
\makeatother
\urlstyle{leo}

\def\pprw{8.5in}
\def\pprh{11in}

\setlength{\paperwidth}{\pprw}
\setlength{\paperheight}{\pprh}
\setlength{\pdfpagewidth}{\pprw}
\setlength{\pdfpageheight}{\pprh}

\definecolor{linkColor}{RGB}{6,125,233}
\hypersetup{%
  bookmarksnumbered,
  colorlinks,
  citecolor=black,
  filecolor=black,
  linkcolor=black,
  urlcolor=linkColor,
  breaklinks=true,
}

\WithSuffix\newcommand\caption*{\caption}

\newcommand{\leveldown}
  {\let\section\subsection%
   \let\subsection\subsubsection%
   \let\subsubsection\paragraph%
   \let\paragraph\subparagraph%
  }
  
\newcommand{\levelup}
  {\let\subparagraph\paragraph%
   \let\paragraph\subsubsection%
   \let\subsubsection\subsection%
   \let\subsection\section%
  }

\section{Introduction}\label{introduction}

Seeing the rise and ubiquitousness of the digital devices in recent
years, many Human-Computer Interaction (HCI) researchers, designers and
technologists have started looking at computers not merely as devices to
interact with, but as a support of positive changes in their users'
lives. Examples of this can be seen in the recent combination of
positive psychology \citep{Seligman2000} and the fields of design
\citep{Desmet2012} and HCI \citep{Calvo2014}. There are a multitude of
factors that have been shown to contribute to subjective well-being and
Calvo and Peters \citep{Calvo2014} have made an extensive inventory of
them. In this paper, we present different interactive systems that we
created to foster introspection -- i.e.~to reflect on our own selves.
More specifically, we focused on \emph{interoception} -- the ability to
sense one's internal bodily signals \citep{Farb2015} --,
\emph{contemplation} and ultimately \emph{mindfulness} -- a broad and
nonjudgmental present-moment awareness.

Traditional HCI is often focused on manipulating digital information
through devices. The vision of Ubiquitous Computing, as proposed by
Weiser \citep{Weiser1991}, instead envisioned digital technology fused
in the environment itself, only intervening when necessary. One of the
underlying goal was to create \emph{calm} experiences
\citep{Weiser1996}. We present systems that are rooted in this
philosophy of infusing the real world with a digital overlay so as to
keep user's awareness of the environment and create such experiences.
For this purpose, we are mostly leveraging tools from Spatial Augmented
Reality \citep{Raskar1998a} and Tangible User Interfaces
\citep{Ishii1997}.

Notably, for interoception, we proposed two different augmented tangible
avatars (Figure \ref{teaser}, left and middle) representing real-time
physiological measures -- ElectroEncephaloGraphy (EEG),
ElectroCardioGraphy (ECG), ElectroDermal Activity (EDA), breathing.
Also, we developed tools to craft tailored representations of such
signals. Then, based on the toolkit we created, we target contemplative
and mindfulness practices. We built a first prototype of an augmented
physical toy that took the form of an augmented sandbox hosting a
digital world, which properties are mapped to the user's breathing and
heart rate (Figure \ref{teaser}, right).

\section{Exposing Inner States}\label{exposing-inner-states}

We developed tools to let people discover and learn about their inner
selves. We got interested in tangible interaction, notably because of
pedagogical principles; learners better become conscious of complex
phenomena through self-investigation and manipulations
\citep{Vosniadou2001, Fleck2013}. Hence, we have been using tangible
augmented objects as a medium to grasp the complexity of physiological
activity and human minds.

Teegi \citep{Frey2014b} is a ``puppet'' that displays the users' brain
activity in real time -- recorded by the mean of EEG. Teegi proved to be
an appealing form factor, that favored the understanding of how the
brain works, even for novices with little or no prior knowledge. The
fact that Teegi is anthropomorphic encourages users to ``connect'' with
their avatars, as the combination of anthropomorphism and tangibility
foster social presence and likability
\citep{Schmitz2010, Hornecker2011, Horn2009}. Moreover, this bond helps
them comprehend rather abstract information -- i.e.~brain activity --
that they could not directly sense otherwise. Indeed, cognitive
processes are difficult to assimilate \citep{Nisbett1977}.

\begin{figure*}
  \centering
  \subfloat[\label{fig:sensors-ecg}]{\includegraphics[width=0.240\hsize]{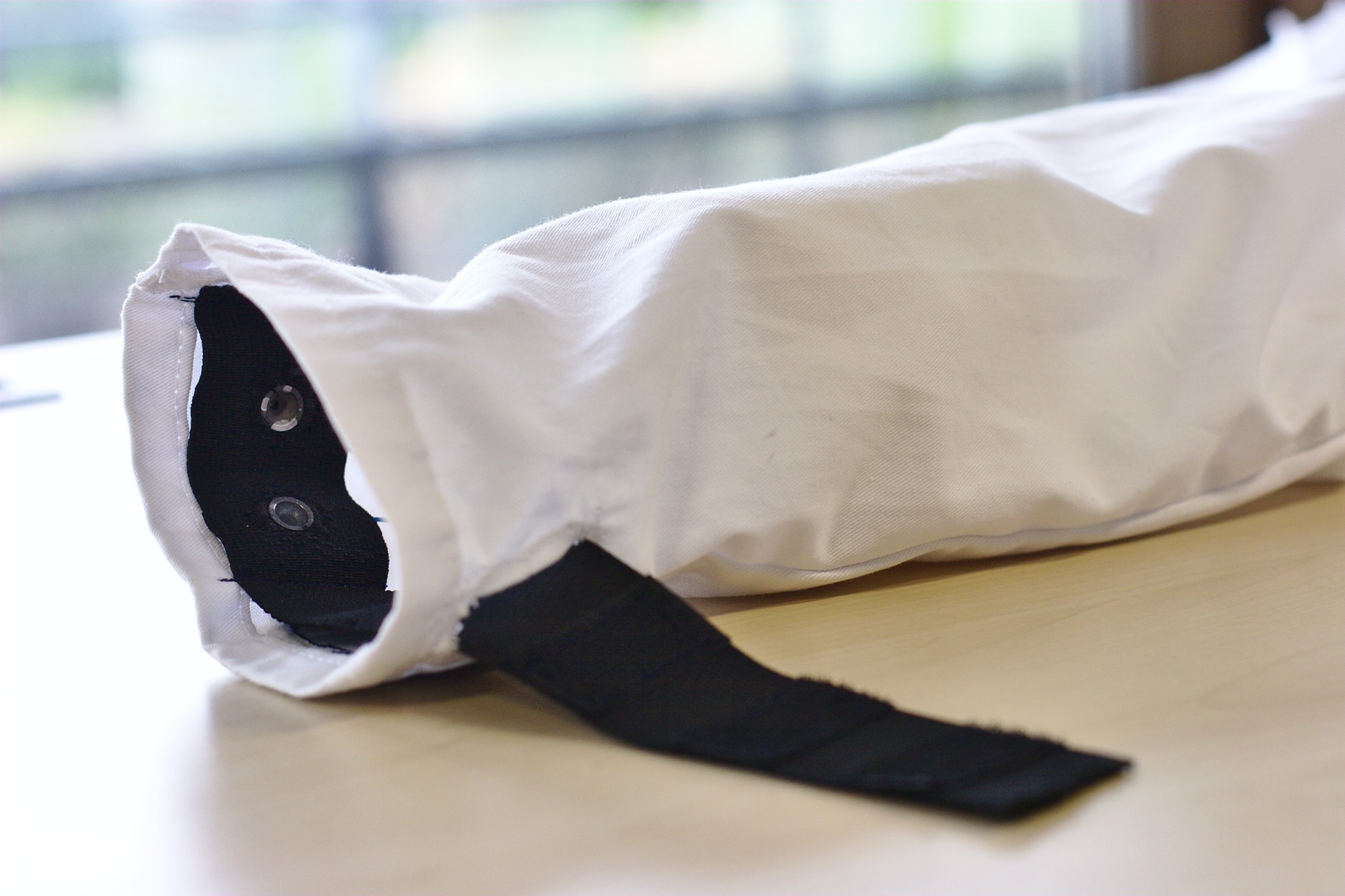}} \hfill
  \subfloat[\label{fig:sensors-eda}]{\includegraphics[width=0.240\hsize]{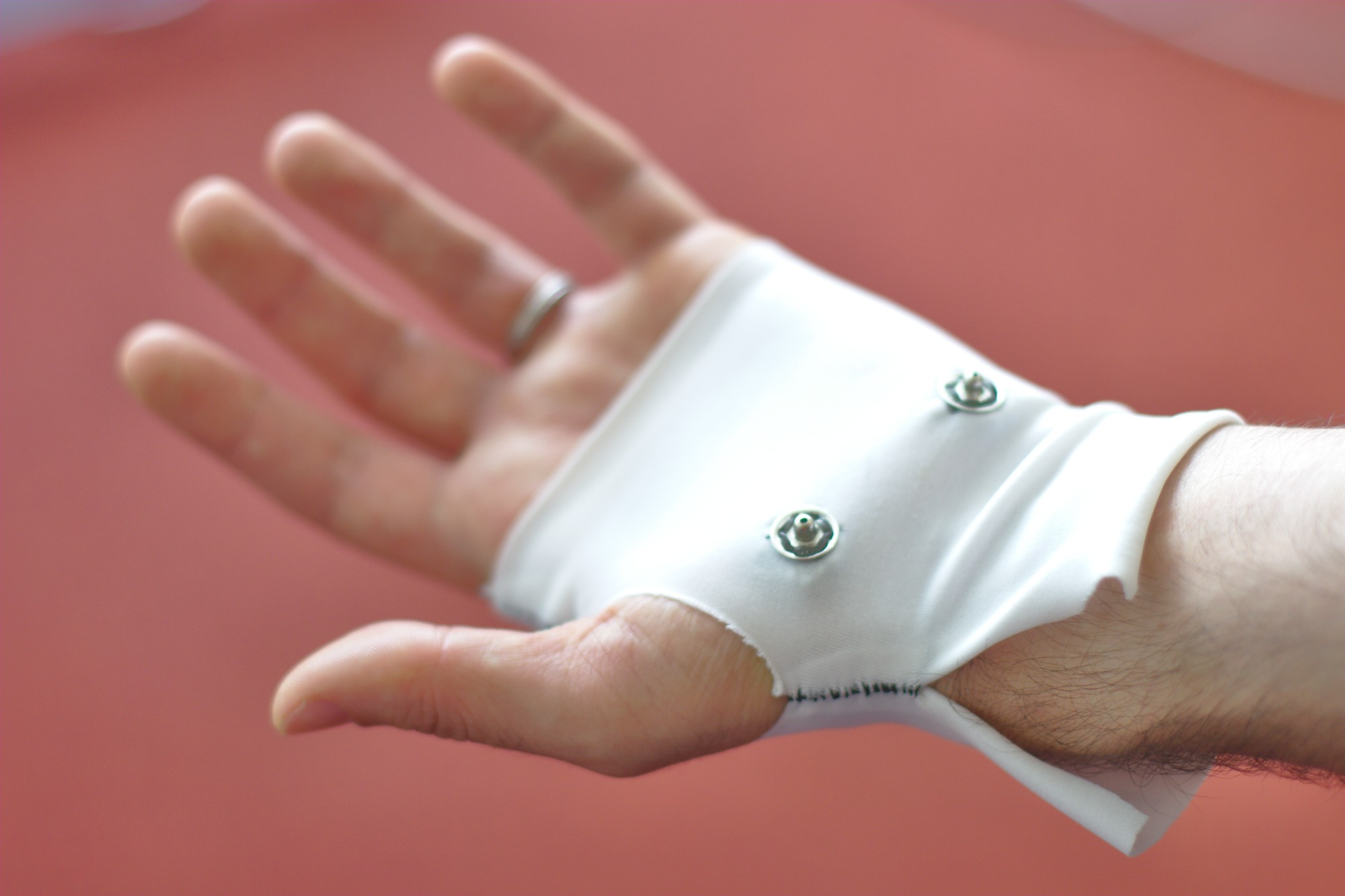}} \hfill
  \subfloat[\label{fig:sensors-breathing}]{\includegraphics[width=0.240\hsize]{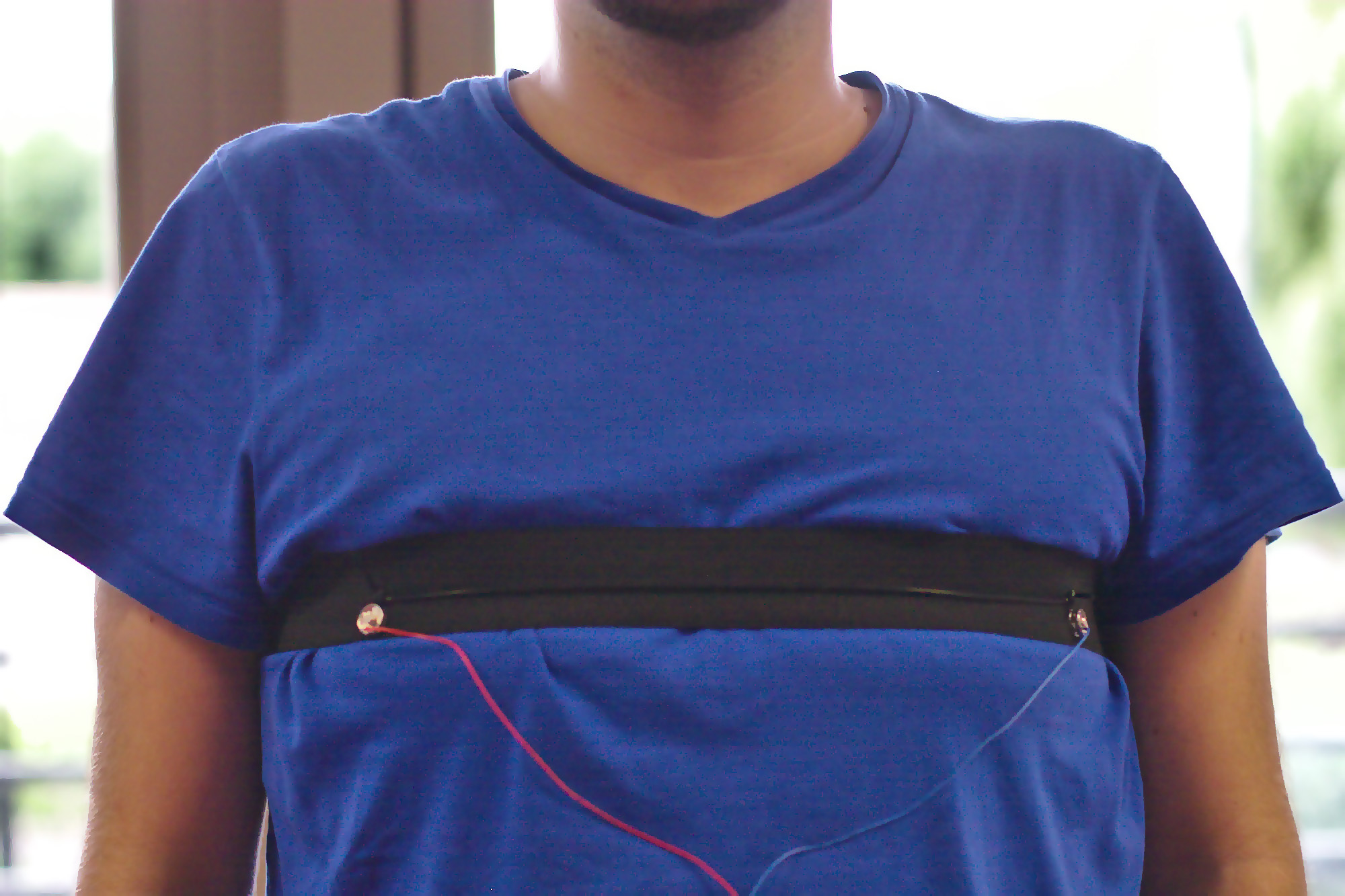}} \hfill
  \subfloat[\label{fig:sensors-openbci}]{\includegraphics[width=0.240\hsize]{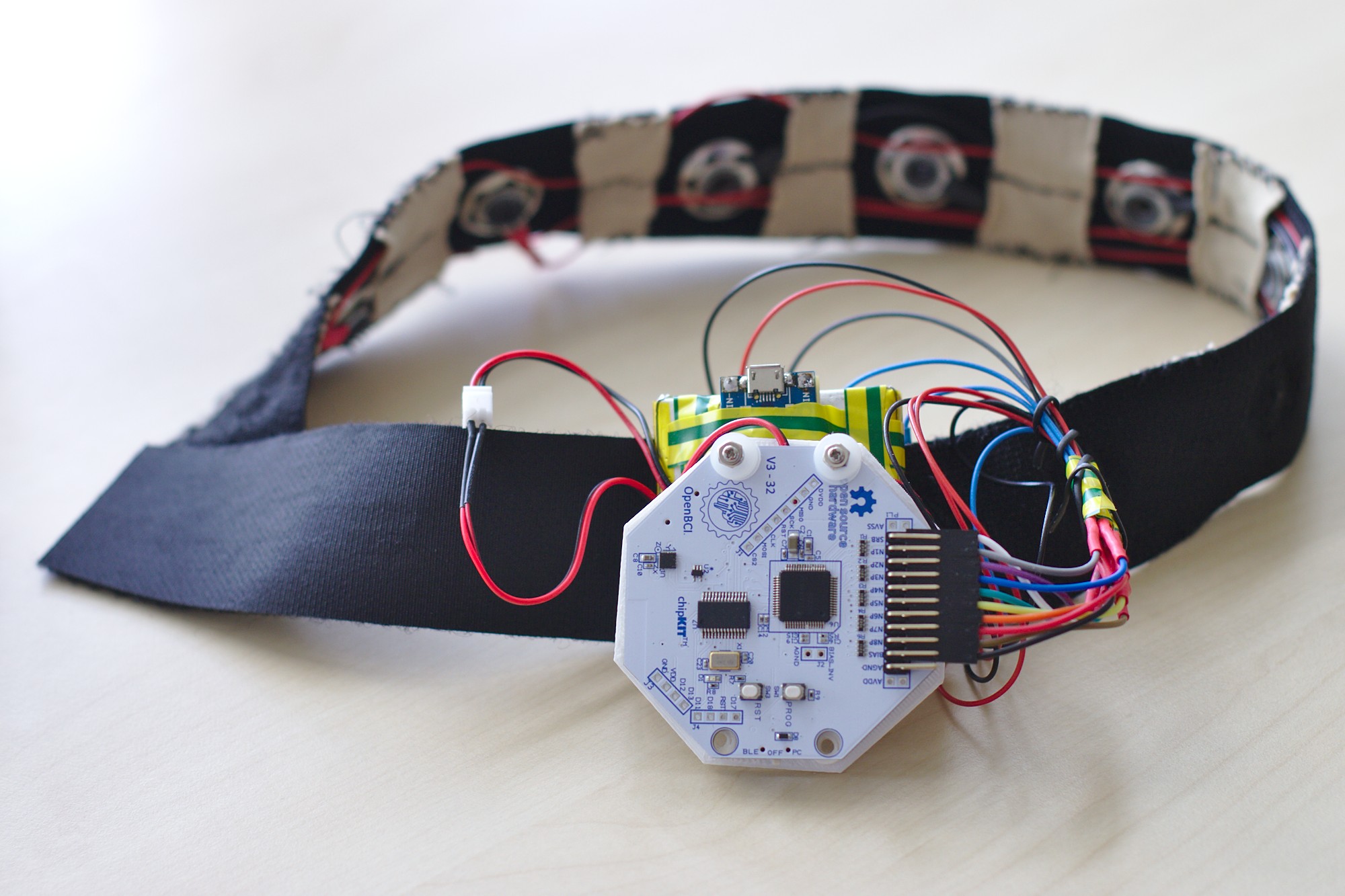}}
  \caption{Wearables. a: coat embedding ECG sensors; b: fingerless glove
measuring EDA; c: breathing belt; d: EEG headband.}\label{fig:sensors}
\end{figure*}

Teegi was a dedicated support for visualizing brain activity but did not
propose any high-level constructs such as workload or attention. We went
beyond this interface with the Tobe project \citep{Gervais2016}, which
gave access to mental states alongside low level signals
(e.g.~heartbeats, breathing). Moreover, we designed wearable sensors,
that could be equipped easily (Figure \ref{fig:sensors}).

Because the dynamic representation of physiological signals is still an
open question at the moment \citep{Chanel2015}, we conducted surveys
during a pilot study to gain more insight about the knowledge and the
representation people had. In particular, we asked participants to
express with drawings and text how they would represent various metrics.
We were surprised by how creative people were; we found little
resemblance between participants for a given signal, even where we would
have expected a consensus to emerge -- e.g.~some people drew a
physiologically accurate heart instead of a simple sketch. Therefore,
the toolkit we built around Tobe also enables people to tailor their own
dynamic feedback, animated using their real-time physiological
measurements.

Not only does customized bio- and neurofeedbacks create meaningful
representations, but the very act of taking part in the creation of the
avatar, through 3D printing, increases the involvement of users in the
process. But far from gaining simple knowledge, such emphasis on the
body has immediate benefits. For instance, improving interoception could
suffice to favor a better health \citep{Farb2015}. Since interoception
is also associated with empathy, it could affect how we feel toward
others \citep{Fukushima2011}. By exposing overtly inner states -- as we
propose to do with the tools we present -- we sharpen human senses
toward themselves and others, for the better good of both.

\section{Supporting Mindfulness}\label{supporting-mindfulness}

One of our overarching goals was to eventually go beyond ``simple''
feedback and interoception, and also directly support mindfulness
activities.

As a first step in this direction, we put into practice the Tobe system
to create a multiuser relaxation experience, where pairs of users had to
synchronize their heart rates through breathing exercises (Figure
\ref{teaser}, left). From there, it encouraged us to go further and to
build a dedicated application and form factor for mindful practices.

Inner Garden \citep{Roo2016} is an augmented sandbox: a world in
miniature that slowly evolves according to the users' inner state. The
system is inspired by both the reflective and metaphoric nature of zen
gardens as well as the playful and experimental nature of sandboxes. Zen
gardens are all about careful placement of elements and are often used
for contemplative and meditative purposes. On the other hand, sandboxes
call for interaction and experimentation. Our main goal was to create an
ambient and meditative toy that could both include playful physical
creation and contemplation.

We designed Inner Garden as a toy: it can be played with, without any
inherent goal or rules. Since our objective was to create a
self-reflective and slow experience, mainly driven by self-motivation
and curiosity, a toy seemed the best support. Using this approach,
Paulos et al. \citep{Paulos2014} created toys to encourage children to
explore their physical environment. Another example is the work of
Karlesky et al. \citep{Karlesky2014} who created seemingly meaningless
tangible toys in order to explore the interaction that happens in the
margins of creative work. The sandbox itself can be made of any size. We
built one that is small enough to live on the side of the desk, acting
as an ambient display during the user's daily activities, while the user
can also decide to interact directly. Alternatively, a bigger one could
support more users in a shared environment, such as a living room at
home or a break area at work.

Whenever the user alters the terrain, the modified section of the
simulation is restarted, living on its own. Grass grows, trees starts
appearing and water flows. However, this process takes place slowly. The
growth speed and overall health of the world, in combination with
atmospheric effects (day-night time-lapse, clouds and sea) are linked to
the user's inner state, measured with EEG and a breathing belt.

The main objective of the system is to foster contemplative and
reflective activities. The small living world is a peaceful, slow
display, while also motivating the users to practice relaxing breathing
exercises whenever they want to take a moment for themselves.

\section{Discussion}\label{discussion}

In our undergoing Inner Garden project, we are interested both in social
interactions and in individual user experience. We would like to install
the Inner Garden in shared spaces, such as break rooms or living rooms.
Different colleagues or family members could work together onto shaping
the world, and then the world would evolve according to their states --
monitored using wearable sensors such as smartwatches -- e.g.~Empatica's
E4\footnote{\url{https://www.empatica.com/}}. This could lead to
interpersonal interactions, both in working together to improve the
garden and cultivating empathy between participants.

We envision the inclusion of Virtual Reality (VR) using a Head Mounted
Display (HMD) -- such as the Oculus Rift\footnote{\url{https://www.oculus.com/}}
-- enabling a user to have a private session \emph{inside} the garden.
While immersed, the world evolution would be slowed down, and the user
could experience the biofeedback directly (e.g.~his or her breath
controlling the sea waves or the wind). Another advantage of the use of
VR is that, since the user is already being equipped with the HMD, the
addition of physiological sensors could be done seamlessly, such as EEG
electrodes embedded in the HMD's elastic strap -- see also
\citep{Hernandez2015} for an example on how to use head mounted devices
to measure breathing and heart rate.

\subsection{What we want to measure}\label{what-we-want-to-measure}

As we propose to make available a shared space that people could
interact with and contemplate, we want to learn how such ambient system
could alter how we relate to one another. We believe that, by itself, an
augmented sandbox left at disposal in a working environment or at home
could prompt direct interactions between passersby, improving
connectedness -- e.g. ``team building''. We envision to test the system
for a prolonged period of time (several weeks), so we could measure
potential changes in interpersonal relationships. Notably, it will be
important to assess if these changes are due to a novelty effect and if
they eventually will settle back to the previous dynamics.

We also plan to use the Inner Garden to propose mindfulness experiences
and familiarize people with introspection. We will let users immerse
themselves in the Inner Garden using a HMD, and we will augment their
sensing capabilities by the mean of physiological sensors. Hence, we are
curious to know if people will seize these technologies as an
opportunity for self-empowerment and we wonder how much the system could
guide them toward healthier habits.

The technology we developed is a mean, not an end. The Inner Garden is a
stand for raising awareness toward others and our own selves. We seek to
measure how the beneficial effects, such as a mindful routine, induced
by the Inner Garden (if any) echoes beyond its direct usage. Ultimately,
it would be desirable that these effects persist months after the system
is removed.

\subsection{How to Measure}\label{how-to-measure}

Inquiries shall be the principal evaluation method to gather insights
about how people welcome the system. Beside surveys, we plan to conduct
regular interviews; freely discussing with several users will notably
help to assess the social impact of the system and gauge which aspects
of the Inner Garden should be refined.

However, the very hardware that we employ to run the system is an
opportunity to also collect exocentric measures, as defined in
\citep{Frey2014a}. Indeed, the cameras that we use for augmenting the
sandbox can also record how users interact with the system and in the
surrounding space. Such video feed could then be processed to observe
peoples' behavior -- e.g.~number and duration of the interactions, if
they rather play with the sandbox or use the virtual reality modality,
to what or whom they look at, how much they move, and so on.

Moreover, the physiological signals that are used to give an online
biofeedback to users could be recorded and further analyzed offline.
Features such as breathing patterns, heart rate variability or
synchronization between brain areas could give metrics about relaxation
or meditative state -- see \citep{Gervais2016}. Such metrics could be
investigated over weeks to sense how users evolve.

Finally, being able to bring altogether inquiries, behavioral data and
physiological sensors -- varying from qualitative to quantitative
measures -- will help to contextualize our findings so we could iterate
over our system, aiming to support mindfulness practices, especially
mindfulness meditation sessions.

\section{Conclusion}\label{conclusion}

We have presented three different interactive systems based on tangible
interaction to foster introspection. More specifically, we described how
anthropomorphic avatars could raise awareness on the inner working of
our bodies and minds and support interoception. We also presented a
first prototype of another form factor of our toolkit, an augmented
sandbox specifically dedicated to promoting mindfulness activities. We
believe that tangible ambient installations, as opposed to regular
displays, have great potential for these purposes since they are
anchored in reality, which mindfulness is all about.

\balance{}

\bibliography{introspectibles}

\end{document}